\def\about {$\sim$}
\def\pks {PKS\,1246{\tt -}410}
\def\solmass {M$_\odot$}
\def\peryr {yr$^{-1}$}
\def\HI {H\kern0.1em{\sc i}} 
\def\radm {rad m$^{-2}$} 
\def\etal   {{\sl et~al.\ }}

\documentstyle[psfig]{mn}

\begin{document}
\title{Magnetic Fields in the Centaurus Cluster}
\author[]
{\parbox[]{6.in} {G. B. Taylor$^1$, A. C. Fabian$^2$ and S. W. Allen$^2$\\
\footnotesize
1. National Radio Astronomy Observatory, Socorro, NM 87801, USA,
gtaylor@nrao.edu \\ 2. Institute of Astronomy, Madingley Road,
Cambridge CB3 0HA, acf@ast.cam.ac.uk; swa@ast.cam.ac.uk}}

\maketitle
\begin{abstract}
We present multi-frequency VLA observations of the radio galaxy
PKS\,1246{\tt -}410 at the center of the Centaurus Cluster, and
compare these results to recent Chandra observations.  The unusual
radio morphology of \pks\ probably results from interactions with the hot,
X-ray emitting gas.  This gas, along with cluster magnetic fields,
also produces substantial Faraday Rotation Measures towards \pks.  We
discuss these observations in the context of a sample of 14 luminous
X-ray clusters with embedded radio galaxies and possible cooling
flows.  
A correlation is found between the
cooling flow rate and the maximum Faraday Rotation Measures.  Magnetic
fields of strength 10--40 $\mu$G are found to be common to the centers
of clusters with strong cooling flows, and somewhat lower field
strengths of 2--10 $\mu$G are found in the non cooling-flow clusters.

\end{abstract}


\section{Introduction}

Radio sources embedded within rich clusters of galaxies can serve as
useful probes of the intracluster medium (ICM). The thermal ICM
pressure may confine the radio source and polarized radiation from
embedded radio sources may also be rotated by the Faraday effect if
magnetic fields are present in the ICM.

The X-ray emission in many clusters is strongly peaked at the center,
indicating high densities and cooling times of the hot ICM in the
inner \about100 kpc which are much less than the Hubble time.  To
maintain hydrostatic equilibrium, either another source of pressure or
an inward flow is required (for a review see Fabian 1994). The cooling
flow phenomenon in clusters is quite common with the short cooling
time condition occurring in 70-90\% of an X-ray flux-limited
sample of clusters (Edge, Stewart \& Fabian 1992; Peres et al 1998).

While most extragalactic radio sources exhibit Faraday rotation
measures (RMs) on the order of 10s of \radm\ due to the interstellar
medium of our galaxy, there are a small number of sources displaying
RMs in excess of \about1000 \radm.  Studies of these high RM
sources (e.g. Cygnus~A: Dreher, Carilli \& Perley 1987; A1795: Ge \&
Owen 1993; Hydra A: Taylor \& Perley 1993) indicate that these RMs are
most likely to be induced by a cluster magnetic field.  
The high RM sources are typically surrounded by luminous, X-ray emitting 
gaseous halos, and in most cases a cooling flow (Taylor, Barton \& Ge 1994).  The diversity of the radio structures
and powers, and the commonality of the X-ray parameters suggest that
the high RMs are the result of external Faraday rotation occurring 
on large scales, and are not due to a local interaction between 
the radio source and its environment. For
external Faraday rotation, the RMs are related to the density, $n_{\rm
  e}$, and magnetic field along the line-of-sight, $B_{\|}$, through
the cluster according to
$$ RM = 812\int\limits_0^L n_{\rm e} B_{\|} {\rm d}l ~{\rm
  radians~m}^{-2}~,
\eqno(1)
$$
where $B_{\|}$ is measured in $\mu$Gauss,  $n_{\rm e}$ 
in cm$^{-3}$ ~and d$l$ in kpc.  The RM distribution, along
with X-ray observations (used to estimate $n_{\rm e}$), can then be
used to understand the magnetic field structure along the
line-of-sight (see review by Carilli \& Taylor 2002).

Here we present VLA (Very Large Array)\footnote{The National Radio Astronomy Observatory
is a facility of the National Science Foundation operated under a
cooperative agreement by Associated Universities, Inc.} observations
of the heretofore little known radio galaxy PKS\,1246{\tt -}410 at the centre 
of the
Centaurus cluster.  This radio galaxy should not to be confused with the
more famous radio galaxy Centaurus~A.  In \S 2 we describe the radio
observations.  In \S 3 we summarize the X-ray observations, and in \S
4 we present our analysis of both the radio and X-ray observations.
In \S 5 we present our current knowledge of the sample of 14 clusters
with embedded radio galaxies originally selected by Taylor \etal
(1994).

We assume $H_0$ = 50 km s$^{-1}$ Mpc$^{-1}$ and $\Omega$
 = 1 throughout.

\section{Radio Observations and Data Reduction}

Observations of PKS\,1246{\tt -}410 were made at 1.465, 1.565, 4.635,
4.885, 8.115, and 8.485 \,GHz at the VLA.  
The details of the observations are provided in Table 1.  The
source 3C\,286 was used as the primary flux density calibrator, and as
an absolute reference for the electric vector polarization angle
(EVPA).  Phase calibration was derived from the nearby compact source
J1316{\tt -}3338 with a cycle time between calibrators of about 12 minutes.
Instrumental calibration of the polarization leakage terms was
obtained using OQ\,208 which was observed over a wide range in
parallactic angle.  The data were reduced in AIPS (Astronomical Image
Processing System), following the standard procedures.  The AIPS task
IMAGR was used with a suitable taper to make $I$, $Q$ and $U$ images at each
of the 6 frequencies observed at the same resolution.  Polarized
intensity, $P$, images and polarization angle, $\chi$, images were
derived from the $Q$ and $U$ images.  The Faraday Rotation Measure (RM)
image was made from the combination of the 4 higher frequency $\chi$ images for PKS\,1246{\tt -}410.
Pixels in the RM image were flagged when the error in any of the 
$\chi$ images exceeded 15 degrees.
Stokes $I$ images using multiple
frequencies within an observing band were also produced. In addition
to the added sensitivity, the image produced also benefits from
improved ($u,v$) coverage.  No correction for the spectral dirty beam
has been performed, but the sidelobes expected from this effect 
should be below the thermal noise floor.

\section{X-ray Observations}

Allen et al.\ (2001a; see also Ikebe et al. 1999) 
present ASCA observations of the Centaurus
cluster.  They find that the spectrum of the central 100 kpc can 
be well-described by either a two-temperature or cooling flow model 
with a mass deposition rate of 30--50 \solmass~\peryr. 

Sanders \& Fabian (2001) present data from Chandra observations at
much higher spatial resolution and report the 
discovery of a knotty, plume-like feature that begins 
60\arcsec\ (17
kpc) to the northeast of the nucleus, and curves to the south before
terminating within 1\arcsec\ (0.28 kpc) of the radio core (see \S
4.2). Sanders \& Fabian find that this plume is 0.25 keV cooler than
the surrounding thermal gas and report on the presence of a ``cold
front'' in the Centaurus cluster where the temperature sharply
increases near the end of the plume, to the north-east of the core. One
way to produce such features is by suppression of thermal conduction by
magnetic fields (see Carilli \& Taylor 2002 and references therein).

The electron density profile for the central 70 kpc ($\sim$
4.2\arcmin) radius region, determined from a deprojection analysis, is
shown in Fig.~1.  The best-fitting $\beta-$model for the density has
parameter values $n_0 = 0.09\pm0.004$ cm$^{-3}$, $\beta =
0.62\pm0.02$, and a core radius, $r_c = 5.66\pm0.04$ kpc. Note that
the core size is similar to that of the radio source, which has made
`holes' in the X-ray emission at radii of 1--5~kpc.

\section{Results}

\subsection{Radio Properties of \pks}
  
The radio source PKS\,1246{\tt -}410 is associated with NGC~4696, an
elliptical galaxy at the center of the Centaurus cluster (also known
as Abell 3526).  The redshift is 0.0099, so 1\arcsec\ corresponds to
0.28 kpc.  The radio power of PKS\,1246{\tt -}410 is 1.5 $\times
10^{24}$ W Hz$^{-1}$ at 1.6 GHz, identifying it as a low power
radio galaxy for which we would expect an FR~1 morphology (edge dimmed
``woofly'' radio tails).  In fact the radio morphology at 1.6 GHz,
shown in Fig.~2, appears quite different from most low power radio galaxies.
The usual well defined core and thin jets feeding into larger lobes
are missing.  Instead there is a bright point source which we identify
as the core based on its central location and slightly flatter
spectral index of $\alpha = -0.5$, where $S_\nu \propto \nu^\alpha$
(Fig.~3).  The position of the core and a summary of source properties
is given in Table~2.  The core is surrounded by ``lobe'' emission with
a predominantly east-west orientation.  At their extremes both lobes
appear (at least in projection) bent to the south and back in towards
the core, although there is also a weak extension to the northeast.
The emission in the extremities is quite steep ($\alpha \sim -2$).
Overall the impression is of a rather distorted radio morphology,
perhaps due to confinement by the hot X-ray emitting gas.  Similar
distortions have been seen for other radio galaxies embedded in dense
X-ray environments with the most extreme case being the amorphous
radio galaxy PKS\,0745{\tt -}191 (Taylor et al.\ 1994). The
swept-back appearance of the outer parts of the lobes may be due to
relative motion of the galaxy and ICM (see Sanders \& Fabian 2001 for
more discussion on this point).

Fortunately, and unlike PKS\,0745{\tt -}191, the bright radio lobes of
PKS\,1246{\tt -}410 are moderately well polarized (5 -- 40\%), so it
is possible to derive the RM distribution across nearly the entire
radio source.  This has been done using four frequencies spread across
the 4 and 6 cm bands, and is shown in Fig.~4.  The RMs appear patchy
and are not correlated with any features in total intensity.  The RMs
in PKS\,1246{\tt -}410 are typically in the range $-500$ to $+500$
\radm\ and reach a maximum of 1800 \radm.  A histogram of the RM
distribution is shown in Fig.~5.  After correcting for Faraday
Rotation, the projected intrinsic magnetic field orientation of the
synchrotron emission generally follows the edges of the source
(Fig.~6).  This is typical for extended radio sources.

\subsection{X-ray/Radio Comparisons}

In Figs.~7 and 8 we show overlays of the X-ray and radio emission.
This cluster shows a dramatic plume in the X-rays that reaches 
in to the cluster center and wraps around the radio galaxy.  At
the highest spatial resolution the radio emission seems to emerge
from the core along a nearly N-S axis (Fig.~9).  The rapid change to a 
more nearly E-W orientation of the radio emission is most likely
the result of interactions with the thermal gas.  Further from
the nucleus at $\sim$3 kpc the radio lobes bend to the south.
The eastern lobe appears to fill in the region bounded by the plume,
while the western lobe also spills around it.  In general the 
impression is that the radio source has been contained by the 
dense gas in 
the plume.  Similar examples of radio emission filling X-ray
cavities and avoiding overdensities have been
found for 3C\,84 at the center of the Perseus cluster 
(Bohringer et al.\ 1993, Fabian et al.\ 2000), in PKS\,1508+059 in
A2029 (Taylor et al.\ 1994) and in Hydra A in A780 
(McNamara et al.\ 2000).

\subsection{Magnetic Field Estimates}

The gas density profile derived from ROSAT HRI observations
can be parameterized with a $\beta$-model fit as described
in \S 3.
This gives an electron density at radius 10 kpc of 
0.04 cm$^{-3}$.  Considering the patch of $-$1500 \radm\ located
3 kpc west of the core, the minimum constant magnetic field, $B_{\|}$, that 
could produce the observed RM over a path length of 10 kpc has a strength of 8 $\mu$G.
Several changes in sign are in evidence
on both sides of the core, however, indicating that the magnetic field
direction is rather chaotic.  The simplest explanation for the RM
distribution is a tangled magnetic field with a coherence length
of \about 1 kpc.  For
a density distribution that follows a $\beta$-profile, Felten (1996)
has derived the following relation for the RM dispersion:

$$
\sigma_{\rm RM} = {{{\rm K} B~n_0~r_c^{1/2} l^{1/2}}
\over{(1 + r^2/r_c^2)^{(6\beta - 1)/4}}}
\sqrt{{\Gamma(3\beta - 0.5)}\over{{\Gamma(3\beta)}}}
\eqno(2)
$$
where $n_0$, $r_c$, and $\beta$ are derived in \S3.2, $l$ is the cell
size, $r$ is the distance of the radio source from the cluster center,
$\Gamma$ is the Gamma function, and K is a factor which depends on the
location of the radio source along the line-of-sight through the
cluster: K = 624 if the source is beyond the cluster, and K = 441 if
the source is halfway through the cluster.  Note that Eq.~2 
assumes that the magnetic
field strength, $B$, is related to the component along the
line of sight, ($B_{\|}$), by $B = \sqrt{3} B_\|$.   Given the clear
interactions between the X-ray and radio plasma, we can assume that
\pks\ is close to the true center of the cluster, and not fortuitously
projected on the center. With the above measurements or assumed
values, we use the estimated RM dispersion of 660 \radm\ from Fig.~5
to estimate the cluster magnetic field strength at 11 $\mu$G.  In \S5.2
we compare this value to those for a larger sample of clusters.

Fig. 8 shows how the X-ray emission varies across the radio lobes. It
is interesting to compare it with Fig.~4 which shows the rotation
measure map. The parts with positive RM (blue in Fig. 4) near the
centre have the highest X-ray surface brightness, $S_{\rm X}$ (brown
in Fig. 8); the parts with negative RM (yellow/orange in Fig.~ 4) have
a lower $S_{\rm X}$ (green in Fig 8) and the parts with little RM
(green in Fig. 4) have the lowest $S_{\rm X}$ (blue in Fig. 8). There
is thus tentative evidence that the regions of high $|RM|$ have high
$S_{\rm X}$ and at least some of the excess X-ray emission lies in
front of the radio source. Since $|RM| \propto n B \ell$ and the X-ray
surface brightness $S_{\rm X}\propto n^2 T^{1/2} \ell$ (the dependence
on temperature $T$ is actually complicated by line emission so the
proportionality is approximate), if the magnetic field $B
\propto \sqrt{nT}$, i.e.  the magnetic pressure is in equipartition
with (or a fixed fraction of) the thermal pressure, then $|RM|\propto
\delta S_{\rm X}n^{-1/2}$, where $\delta S_{\rm X}$ is that part of
$S_{\rm X}$ which is in front of the radio source. Qualitatively this
explains what is seen, but the detailed relationship must depend upon
the field configuration, which must be complex.

\begin{figure}
\centerline{\psfig{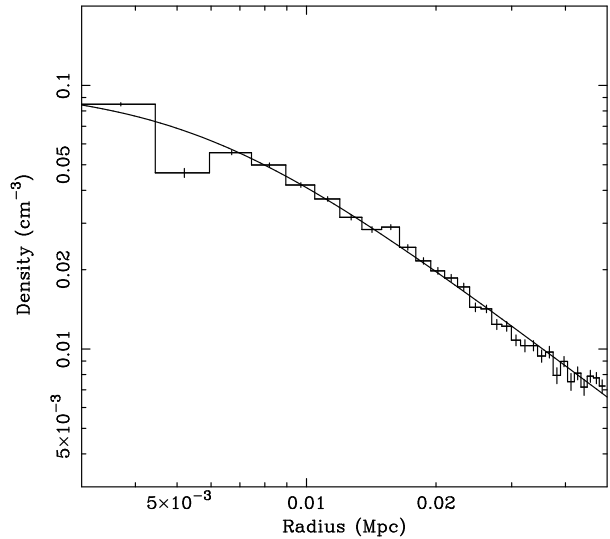}}
\caption{Deprojected density profile determined from the 
Chandra observations. The solid line shows the best-fitting
$\beta-$model. The inner few kpc are affected by the 'hole' in X-ray
emission associated with the radio lobes.
\label{fig1}}
\end{figure}

\begin{figure}
\centerline{\psfig{figure=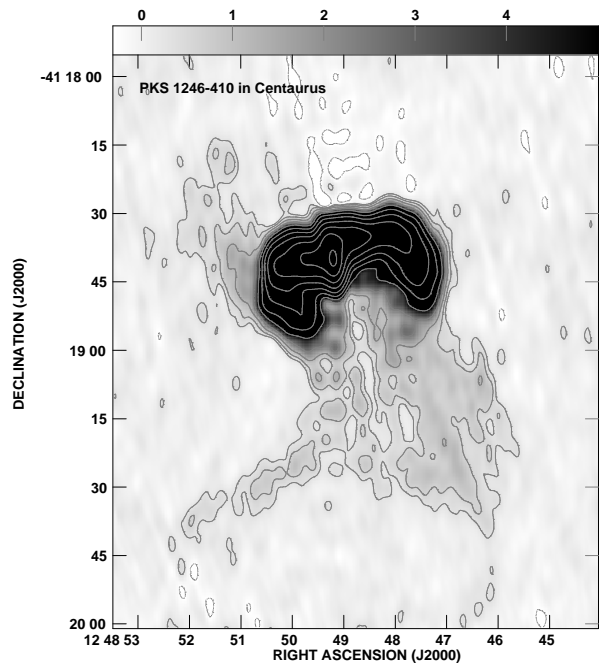,width=0.45\textwidth}}
\caption{VLA image of PKS\,1246{\tt -}410 at 1.565 GHz.  The greyscale 
has been chosen to show the faint extended emission in the range 
0 to 5 mJy/beam.  Contours start at 0.3 mJy/beam and increase by 
factors of 2.  The peak flux density in the map is 358 mJy/beam.
The restoring beam has dimensions 4.4 $\times$ 2.1\arcsec\ in position
angle 0 degrees.
\label{fig2}}
\end{figure}

\begin{figure}
\centerline{\psfig{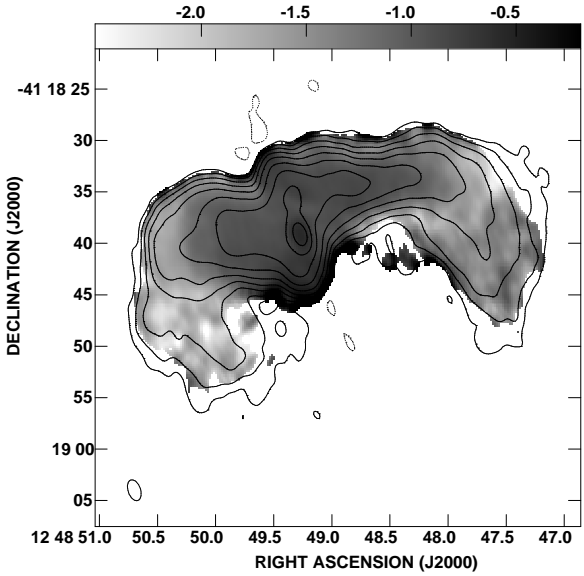}}
\caption{A spectral index map for PKS\,1246{\tt -}410 made by combining
images at 4.76 and 8.30 GHz.
Contours represent the 4.76 GHz total intensity starting at 0.25 mJy/beam
and increasing by factors of 2. The peak in the image is 106
mJy/beam. The restoring beam is plotted in the lower right corner and
has dimensions 2.1 $\times$ 1.2 in p.a. 19$^\circ$.
\label{fig3}}
\end{figure}

\begin{figure}
\centerline{\psfig{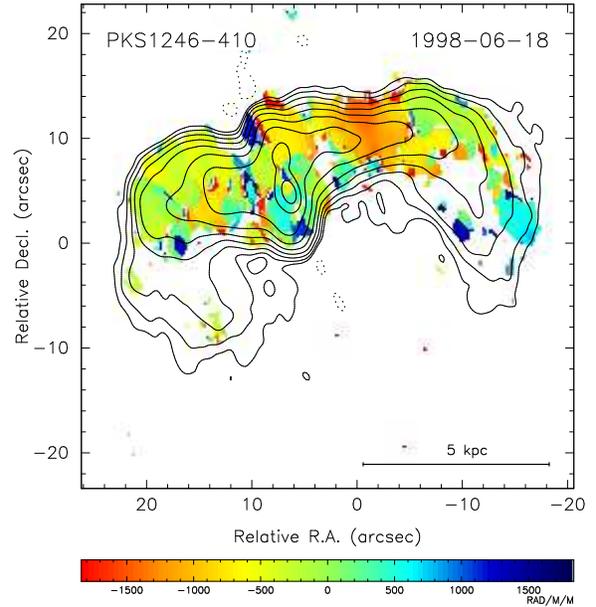} }
\caption{The rotation measure (RM) map for PKS\,1246{\tt -}410 at the
center of the Centaurus cluster. 
Contours are the same as in Fig.~3.
\label{fig4}}
\end{figure}

\begin{figure}
\centerline{\psfig{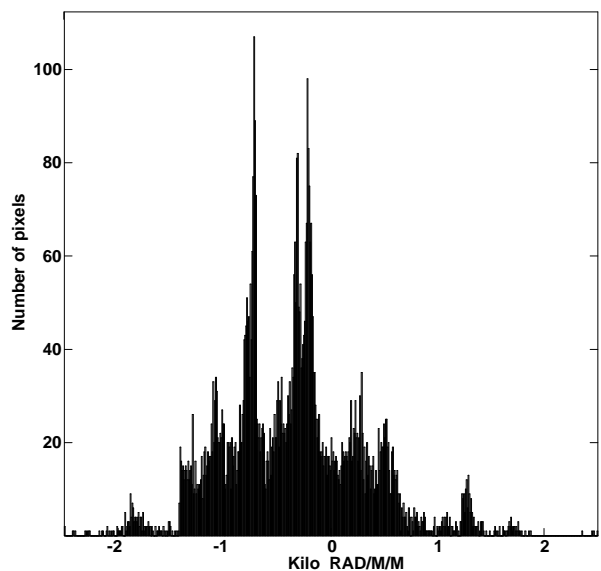} }
\caption{Rotation measure distribution for PKS\,1246{\tt -}410.  The mean value
is $-$356 \radm\ with a dispersion of 662 \radm.  The width of each bin is
8.3 \radm, which is about the same as the error in the RM determination.
\label{fig5}}
\end{figure}

\begin{figure}
\centerline{\psfig{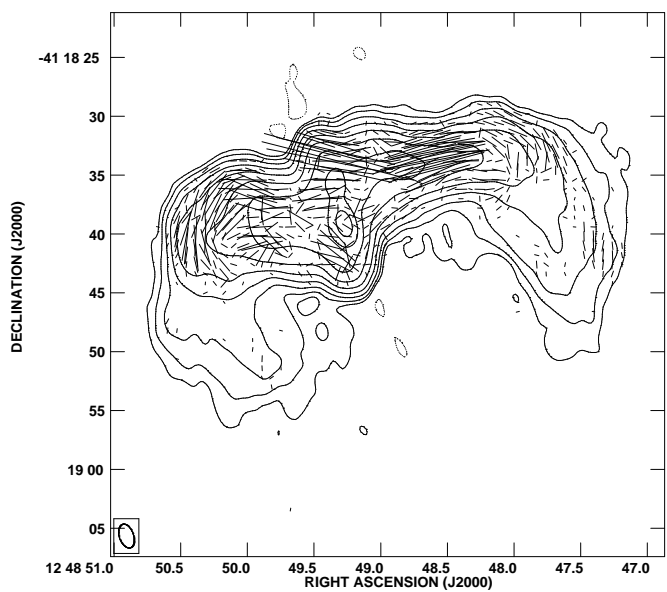}}
\caption{The Projected magnetic field vectors at 
4.9 GHz (vectors corrected for Faraday rotation) overlaid on
total intensity contours for PKS\,1246{\tt -}410.   Contours are the same
as in Fig.~3.  The length of the vectors is proportional to polarized
intensity (1\arcsec\ = 0.31 mJy/beam).
\label{fig6}}
\end{figure}

\begin{figure}
\centerline{\psfig{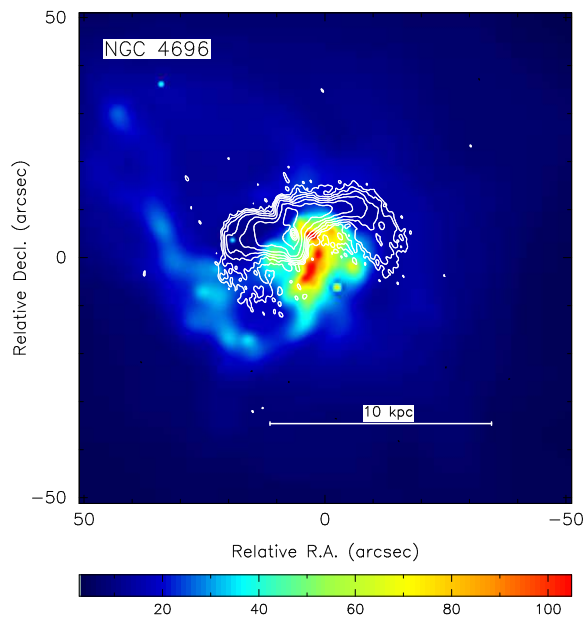}}
\caption{The Centaurus Cluster in the X-ray (false colour from the
Chandra image) and radio (8.4 GHz contours).  The
resolution is similar in both images.
\label{fig7}}
\end{figure}

\begin{figure}
\centerline{\psfig{figure=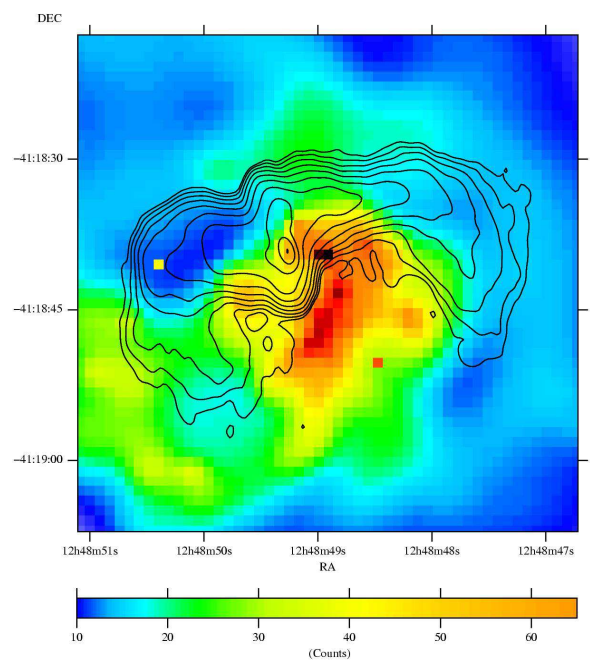,width=0.45\textwidth}} 
\caption{The inner part of the Centaurus Cluster in the X-ray (ct
arcsec$^{-1}$ 31.5 ks$^{-1}$ (0.5--5~keV); false colour from the
adaptively-smoothed Chandra image) and radio (5 GHz contours).
Contours are the same as in Fig.~3.
\label{fig8}}
\end{figure}

\begin{figure}
{\psfig{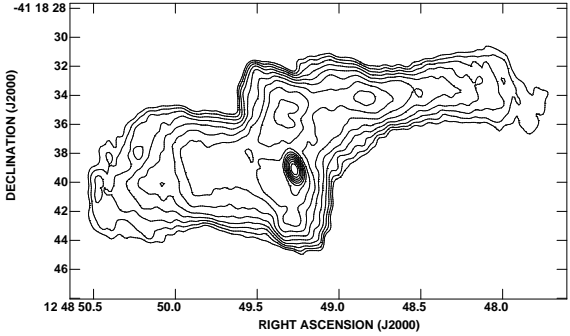} }
\caption{VLA image of PKS\,1246{\tt -}410 at 8.30 GHz made from the BnA
configuration data.  Contours start at 0.35 mJy/beam and increase by 
factors of $\sqrt{2}$.  The peak flux density in the map is 63 mJy/beam.
The restoring beam has dimensions 1.05 $\times$ 0.58\arcsec\ and a 
position angle of 21 degrees.
\label{fig9}}
\end{figure}

\begin{figure}
\centerline{\psfig{figure=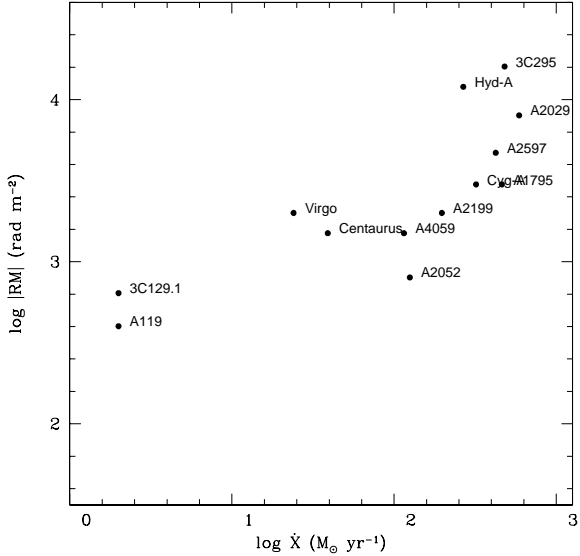,width=0.45\textwidth} }
\caption{The maximum absolute RM plotted as a function of 
$\dot{X}$ for the 13 clusters in Table 3 with measured RMs.  We
also include the well-studied, but higher redshift 3C\,295 cluster 
on this plot. 
\label{fig10}}
\end{figure}

\def\dg{$^{\circ}$}

\begin{table}
\scriptsize
\caption{O{\sc bservational} P{\sc arameters}}
\begin{tabular}{llrrrr}
\hline
\hline
Source & Date & Frequency & Bandwidth$^a$ & Config. & Duration \\
 &  & (GHz) & (MHz) &  & (hours) \\
\hline
PKS\,1246   & Apr1998 & 1465/1565 & 25 & A & 0.24 \\
{\tt -}410         & Jun1998 & 1465/1565 & 25 & BnA & 0.99 \\
         & Apr1998 & 4635/4885 & 50 & A & 0.09 \\
         & Jun1998 & 4635/4885 & 50 & BnA & 1.23 \\
         & Nov1998 & 4635/4885 & 50 & CnB & 0.61 \\
         & Jun1998 & 8115/8485 & 50 & BnA & 0.89 \\
         & Nov1998 & 8115/8485 & 50 & CnB & 0.62 \\
\hline
\end{tabular}
\end{table}

\begin{table}
\caption{S{\sc ource} P{\sc roperties}}
\begin{tabular}{l c c}
\hline
\hline
Property & PKS\,1246{\tt -}410 \\
\hline
core RA (J2000) & 12$^h$48$^m$49\rlap{$^s$}{.\,}273  \\
core Dec. (J2000) &  $-41^o$ 18\arcmin 39\rlap{\arcsec}{.\,}11 \\
velocity & 2958 $\pm$ 15 km s$^{-1}$ \\
projected distance  &  0 Mpc \\
angular size & 122\arcsec \\
physical size & 34 kpc \\
flux density (5 GHz) & 1330 mJy \\
power (5 GHz) & 5.7 $\times$ 10$^{23}$ W Hz$^{-1}$  \\
$<$RM$>$ & $-$356 \radm \\
$\sigma_{\rm RM}$ &  662 \radm \\
$|$RM$_{\rm max}|$ & 1800 \radm \\
\hline
\end{tabular}
\end{table}

\begin{table*}
      \caption{Radio Sources in Cooling Flow Clusters}
         \label{KapSou}
      \[
         \begin{array}{llrrrrrrrrr}
            \hline
            \noalign{\smallskip}
{\rm Cluster~~~~~} & {\rm Radio~~~~~~~~} & z & \dot{X}^a & {\rm log}\ L_{\rm x} 
& {\rm Gal.}^b & S_{5000} & {\rm log}\ P_{5000} & {\rm Size}^c &
RM(1+z)^2 & {\rm Ref.} \\
 & {\rm Source} &  &  ({\rm M}_\odot\ {\rm yr}^{-1}) & ({\rm erg\ s}^{-1}) & {\rm Type} & 
{\rm (Jy)} & ({\rm W\ Hz}^{-1}) & ({\rm kpc}) & ({\rm rad\ m}^{-2}) \\
            \noalign{\smallskip}
            \hline
            \noalign{\smallskip}
{\rm Cygnus\ A} & {\rm 3C~}405  & 0.057 & 320^{+71}_{-32} & 44.8 & E & 364 & 27.7 & 180 & 3000 & 1 \\
{\rm Virgo} & {\rm 3C~}274 & 18\,{\rm Mpc} & 24^{+10.6}_{-5.7} & 43.1 & E & 76 & 24.4 & 58 & 2000 & 2 \\
{\rm A426}       & {\rm 3C~}84 & 0.0183 & 587^{+86}_{-44} & 45.0 & pec & 50 & 25.9 & 640 & N/A & 4 \\
{\rm Hydra\ A}  & {\rm 3C~}218 & 0.0522 & 267^{+48}_{-30} & 44.5 & cD & 14 & 26.3 & 670 & -12000 & 5 \\
{\rm Centaurus} &1246{\tt-}410 & 0.0099 & 39^{+25.5}_{-7.8} & 43.8 & E & 1.4 & 25.8 & 15 & 1500 & 3 \\
{\rm A2052}     & {\rm 3C~}317 & 0.0348 & 125^{+26}_{-6} & 44.1 & cD & 1.0 & 24.7 & 39 & -800 & 6 \\
{\rm A2199}     & {\rm 3C~}338 & 0.031  & 197^{+20}_{-41} & 44.5 & cD & 0.48 & 24.3 & 67 & 2000 & 6 \\
\rm {A119}     & 0053{\tt-}015 & 0.044  & 0^{+2}_{-0} & 44.4 & E & 0.40 & 24.5 & 350 & 400 & 7 \\
               & 0053{\tt-}016 & 0.044  & 0^{+2}_{-0} & 44.4 & E & 0.29 & 24.4 & 300 & 150 & 7 \\
\rm {0745{\tt-}191} &0745{\tt-}191 & 0.1028 & 1038^{+116}_{-68} & 45.4 & cD & 0.39 & 25.3 & 25 & N/A & 8 \\
{\rm A2597}    & 2322{\tt-}123 & 0.0824 & 423^{+91}_{-99} & 44.8 & cD & 0.33 & 25.0 & 10 & 4700 & 9 \\
{\rm A1795}   & {\rm 4C~}26.42 & 0.062  & 462^{+108}_{-56} & 44.9 & cD & 0.26 & 24.6 & 15 & 3000 & 6 \\ 
{\rm 3C~}129  & {\rm 3C~}129.1 & 0.0223 & 0^{+4.2}_{-0} & 44.3 & E & 0.22 & 23.4 & 40 & 640 & 10 \\
                & {\rm 3C~}129 & 0.0208 & 0^{+4.2}_{-0} & 44.3 & E & 2.65 & 24.5 & 500 & 260 & 10 \\
{\rm A4059} &    2354{\tt-}350 & 0.0478 & 115^{+57}_{-37} & 44.3 & cD & 0.11 & 24.0 & 70 & -1500 & 8 \\
{\rm A2029} &    1508{\tt+}059 & 0.0767 & 590^{+27}_{-96} & 45.3 & cD & 0.10 & 24.4 & 80 & -8000 & 8 \\
\noalign{\vskip2pt}
\noalign{\hrule}
\noalign{\vskip2pt}
\noalign{\hrule}
            \noalign{\smallskip}
             \hline
         \end{array}
      \]
\begin{list}{}{}
\item[$^{\rm a}$] {The mass flow rate for a critical cooling time of 1.3 $\times$ 10$^{10}$ from 
Allen \& Fabian (1997), White, Jones \& Forman (1997) and  Peres \etal (1998).} 
\item[$^{\rm b}$] {Galaxy type (E: elliptical; pec: peculiar, or cD: central dominant elliptical)}
\item[$^{\rm c}$]{The largest angular size of the radio source in kpc.} 
\item{References - (1) Dreher et al.\ 1987; (2) Owen et al.
1990; (3) This work;
(4) Burns et al.\ 1992; (5) Taylor \& Perley 1993; (6) Ge \& Owen
1993; (7) Feretti \etal 1999; (8) Taylor, Barton, \& Ge 1994;
(9) Taylor, in prep.; (10) Taylor \etal\ 2001}\\
\end{list}
   \end{table*} 

\section{A sample of radio sources embedded in clusters}

\subsection{Summary of results}

Examination of a flux-limited, all-sky X-ray sample of galaxies put
forth by Edge, Stewart, \& Fabian (1992) reveals that clusters with
non-zero mass flow rates frequently display significant RMs. 
Taylor, Barton \& Ge (1994) formed a sample of 14 cooling flow
clusters each containing at least one embedded radio galaxy 
stronger than 100 mJy at 5 GHz.  All of these sources have now been
extensively studied with the VLA, and in most cases considerably better
X-ray observations are also available.  Here we present an analysis
of the combined results for this sample.

Of the 14 clusters in the sample, 10 display an excess of 800 \radm,
two have RMs between 300 and 700 \radm\ (one is A119 discussed by
Feretti \etal\ 1999; and the other is the 3C~129 cluster), and two
could not be measured (PKS0745{\tt-}191 and the Perseus cluster) due
to a lack of polarized flux. Nominal mass-flow rates, X-ray luminosities
(from 2-10 keV), RMs, and other relevant parameters for these 14
clusters are listed in Table 3.  Recent work has shown that the 
nominal X-ray mass flow rates, calculated under the assumption that 
the cooling flows extend out to the cooling radii in the clusters
(where $t_{\rm cool}=13$ Gyr) are likely to overestimate the true mass deposition 
rates from the cooling flows by a factor of 3 or more. This is required
by the fact that the cooling flows are much younger than 13 Gyr 
(Allen \& Fabian 1997; Allen \etal 2001a) and, importantly, by the 
apparent lack of any gas cooling below 1--2~keV in spectra 
taken with XMM-Newton (Peterson
et al 2001; Tamura et al 2001). For display purposes, we shall 
continue to use the nominal X-ray derived values here, simply as a 
characteristic measure of the amount of cooler gas in the cluster core. 
However, to signify that they are not
complete mass cooling rates (although they compare well with the rate
at which gas cools from the cluster virial temperature to about one
third that value), we denote them as $\dot X$ instead of $\dot M$. 

The RM value given is chosen from the region of largest RM which is
coherent across a beamwidth or more with good polarized
signal-to-noise.  We have also plotted in Fig.~10 the RM as a function
of cooling rate, $\dot{X}$.  These results strongly suggest a
correlation between the cooling rate and rotation measure. Even more
striking is the fact that {\it all} of the known high RM sources (RM
$>$ 700 \radm; see Table 1 in Taylor, Inoue, \& Tabara 1992) at low
redshifts (z $<$ 0.4) are included in this X-ray selected sample.  The
high RM radio galaxy 3C\,295 (Perley \& Taylor 1991) at a redshift of
0.461 is also associated with a luminous X-ray cooling flow (Henry \&
Henriksen 1986; Allen et al.\ 2001b), but the pre-Chandra X-ray data for the
remaining seven high RM, high $z$, sources are not good enough to
distinguish between extended thermal emission and that from an active
nuclear source.  Beyond redshifts of \about 0.4, low power radio
sources like PKS\,1508{\tt+}059, and \pks\ are much too weak to allow
for a determination of their RM with the current generation of radio
telescopes.

\subsection{Cluster Magnetic Field Strengths and Topologies}

Our analysis of the RM dispersion in \pks\ leads to an estimated
cluster magnetic field strength of 11 $\mu$G in the Centaurus cluster.
This is about a factor of two larger than the field strengths of 6
$\mu$G estimated for the two non-cooling flow clusters in our sample
(3C\,129 and A119) by the same method.  At the other end of the range,
applying this method to Hydra~A leads to an estimated field strength
of 35 $\mu$G.  

The scale size of the RM fluctuations in Centaurus is quite small
at $\sim$1 kpc.  In cluster radio sources with both lower and higher RMs 
(e.g. A119, 3C129, Hydra~A) typical scale lengths are $\sim$3 kpc. 

The small physical size of \pks\ only allows us to probe the magnetic
fields along the line-of-sight to the center of the cluster.  Strictly
speaking the region across which the RMs can be measured is only a
projected distance of $\sim$10 kpc, so the RMs could be explained in
terms of the host galaxy.  However, in other clusters such as A119
(Feretti et al. 1999) and A514 (Govoni et al.\ 2001), RM studies have
been performed towards several radio galaxies either in or behind the
cluster and the magnetic fields have been found to prevail to large
radii, $\sim$500 kpc.  In a study of background radio sources seen
through clusters, Clarke, Kronberg \& B\"ohringer (2001) argue
convincingly in favor of cluster-wide magnetic fields distributed
throughout the central $\sim$1 Mpc.

\section{Conclusions}

The radio galaxy \pks\ appears to be strongly influenced by the dense
X-ray emitting cluster gas that surrounds it, especially the plume
discovered by Sanders \& Fabian (2001).  The radio -- X-ray
interaction produces an unusually steep spectrum, small source with
bent lobes.  Combined with a cluster magnetic field, the hot gas also
produces Faraday RMs reaching 1800 \radm.  With some assumptions
about the field topology and distribution, a cluster magnetic field of
$\sim$11 $\mu$G is derived.  This field strength is intermediate
between that of clusters in which there is no cooling-flow
present, and those estimated to have a large cooling-flow.

In the near future the Expanded VLA will provide an order of magnitude
increase in sensitivity over the current VLA.  This should allow 
RM studies towards more than a tenfold increase in the number of
sources within a given cluster.  By analyzing the RM distribution
at many lines of sight through a cluster it should be possible to 
determine how the magnetic field changes with cluster radius and 
density.  

\section{Acknowledgments}
This research has made use of the NASA/IPAC Extragalactic Database (NED)
which is operated by the Jet Propulsion Laboratory, Caltech, under
contract with NASA.  ACF and SWA acknowledge the support of the Royal 
Society.

\end{document}